\documentclass[aps,prd,reprint,amsmath,amssymb,superscriptaddress,floatfix,nofootinbib,showkeys,showpacs,preprintnumbers]{revtex4-2} 

\usepackage[dvipsnames]{xcolor}
\usepackage{hyperref}
\hypersetup{    
  colorlinks      = true,
  linkcolor       = {blue},
  linkbordercolor = {white},
  citecolor       = {blue},
  citebordercolor = {white},
  urlcolor        = {blue},
  urlbordercolor  = {white},
}

\usepackage{amsmath} 
\usepackage{xspace} 
\usepackage{mathtools, tabu}
\usepackage{newtxtext, newtxmath}
\usepackage[T1]{fontenc}
\usepackage{graphicx} 
\usepackage{aas_macros} 

\begin{document}
\newcommand{\comment}[1]{}

\newcommand{\avg}[1]{\left\langle{#1}\right\rangle}
\newcommand{\dd}{{\rm d}} 

\newcommand{\beq}{\begin{equation}}
\newcommand{\eeq}{\end{equation}}
\newcommand{\beqa}{\begin{equation}\begin{aligned}}
\newcommand{\eeqa}{\end{aligned}\end{equation}}
\newcommand{\bit}{\begin{itemize}}
\newcommand{\eit}{\end{itemize}}

\newcommand{\hiGpc}{h^{-1} \rm Gpc}
\newcommand{\hiMpc}{h^{-1} \rm Mpc}
\newcommand{\hikpc}{h^{-1} \rm kpc}
\newcommand{\hiMsun}{h^{-1} M_\odot}
\newcommand{\Msun}{M_\odot}
\newcommand{\kms}{\rm km~s^{-1}}

\newcommand{\OmegaM}{\Omega_{\rm M}}
\newcommand{\OmegaB}{\Omega_{\rm B}}
\newcommand{\ns}{n_{\rm s}}

\newcommand{\Mvir}{M_{\rm vir}}
\newcommand{\Rvir}{R_{\rm vir}}
\newcommand{\Mtwo}{M_{\rm 200m}}
\newcommand{\Rtwo}{R_{\rm 200m}}

\newcommand{\DS}{\Delta\Sigma}
\newcommand{\rp}{r_{\rm p}}
\newcommand{\Sigcrit}{\Sigma_{\rm crit}}
 
\newcommand{\zred}{z_{\rm red}}
\newcommand{\zspec}{z_{\rm spec}}
\newcommand{\zCG}{z_{\rm CG}}

\newcommand{\densityunit}{
h^3 \mathrm{Mpc}^{-3}}

\newcommand{\redmapper}{redMaPPer\xspace} 
\newcommand{\redmagic}{redMaGiC\xspace}

\newcommand{\chisqc}{\chi_{\rm color}^2}

\newcommand{\pproj}{p_{\rm proj}}
\newcommand{\pred}{p_{\rm red}}
\newcommand{\pfit}{p_{\rm fit}}

\newcommand{\pspec}{p_{\rm spec}}
\newcommand{\Dz}{\Delta z}
\newcommand{\Dchi}{\Delta\chi}

\newcommand{\Dzmax}{\Delta z_{\rm max}}
\newcommand{\Rlam}{R_{\lambda}}
\newcommand{\zlam}{z_{\lambda}}
\newcommand{\sz}{\sigma_z}
\newcommand{\zcl}{z_{\rm cl}}

\newcommand{\lamAM}{\lambda_{\rm AM}}

\newcommand{\lamtrue}{\lambda_{\rm true}}
\newcommand{\lamphot}{\lambda_{\rm phot}}
\newcommand{\lamspec}{\lambda_{\rm spec}}

\newcommand{\siglam}{\sigma_\lambda}
\newcommand{\siglnlam}{\sigma_{\ln\lambda}}

\newcommand{\Dza}{\Delta z/(1+z)}

\newcommand{\Ncyl}{N_{\rm cyl}}
\newcommand{\Rcyl}{R_{\rm cyl}}
\newcommand{\fcl}{f_{\rm cl}}
\newcommand{\mucl}{\mu_{\rm cl}}
\newcommand{\sigcl}{\sigma_{\rm cl}}

\newcommand{\Nproj}{\mathcal{N}_{\rm proj}}
\newcommand{\fproj}{f_{\rm proj}}
\newcommand{\muproj}{\mu_{\rm proj}}

\newcommand{\sigproj}{\sigma_{\rm proj}}
\newcommand{\dproj}{d_{\rm proj}}
\newcommand{\qproj}{q_{\rm proj}}

\newcommand{\Ncen}{N_{\rm cen}}
\newcommand{\Nsat}{N_{\rm sat}}
\newcommand{\Mmin}{M_{\rm min}}
\newcommand{\siglogM}{\sigma_{\log M}}

\newcommand{\vx}{{\bf x}}

\defcitealias{Myles21projection}{M21}
\defcitealias{Rozo15RM4}{R15}
\defcitealias{Costanzi19projection}{C19}
\newcommand{\Rozo}{\citetalias{Rozo15RM4}\xspace}
\newcommand{\Myles}{\citetalias{Myles21projection}\xspace}
\newcommand{\Costanzi}{\citetalias{Costanzi19projection}\xspace} 

\title{Optical galaxy cluster mock catalogs with realistic projection effects: Validations with the SDSS clusters}
\author{Andy Lee}
\affiliation{Department of Physics, Boise State University, Boise, ID 83725, USA}
\affiliation{Department of Physics, The Ohio State University, Columbus, OH 43210, USA}
\author{Hao-Yi Wu}
\email[Contact author: ]{hywu@smu.edu}
\affiliation{Department of Physics, Boise State University, Boise, ID 83725, USA}
\affiliation{Department of Physics, Southern Methodist University, Dallas, TX 75205, USA}
\author{Andr\'es N. Salcedo}
\affiliation{Department of Astronomy/Steward Observatory, University of Arizona, Tucson, AZ 85721, USA}
\affiliation{Department of Physics, University of Arizona, Tucson, AZ 85721, USA}
\author{Tomomi Sunayama}
\affiliation{Department of Astronomy/Steward Observatory, University of Arizona, Tucson, AZ 85721, USA}
\affiliation{Department of Physics, University of Arizona, Tucson, AZ 85721, USA}
\author{Matteo Costanzi}
\affiliation{Dipartimento di Fisica - Sezione di Astronomia, Universit\`a di Trieste, 34131 Trieste, Italy \\
INAF-Osservatorio Astronomico di Trieste, 34143 Trieste, Italy\\
IFPU - Institute for Fundamental Physics of the Universe, 34014 Trieste, Italy}
\author{Justin Myles}
\affiliation{Department of Astrophysical Sciences, Princeton University, Princeton, NJ 08544, USA}
\author{Shulei Cao}
\affiliation{Department of Physics, Boise State University, Boise, ID 83725, USA}
\affiliation{Department of Physics, Southern Methodist University, Dallas, TX 75205, USA}
\author{Eduardo Rozo}
\affiliation{Department of Physics, University of Arizona, Tucson, AZ 85721, USA}
\author{Chun-Hao To}
\affiliation{Department of Physics, The Ohio State University, Columbus, OH 43210, USA}
\affiliation{Center for Cosmology and AstroParticle Physics (CCAPP), The Ohio State University, Columbus, OH 43210, USA}
\affiliation{Department of Astronomy, The Ohio State University, Columbus, OH 43210, USA}
\affiliation{Department of Astronomy \& Astrophysics, University of Chicago, Chicago, IL 60637, USA}
\author{David H. Weinberg}
\affiliation{Center for Cosmology and AstroParticle Physics (CCAPP), The Ohio State University, Columbus, OH 43210, USA}
\affiliation{Department of Astronomy, The Ohio State University, Columbus, OH 43210, USA}
\author{Lei Yang}
\affiliation{Department of Physics, Boise State University, Boise, ID 83725, USA}
\affiliation{Department of Physics, Southern Methodist University, Dallas, TX 75205, USA}
\author{Conghao Zhou}
\affiliation{Physics Department, University of California, Santa Cruz, CA 95064, USA\\
Santa Cruz Institute for Particle Physics, Santa Cruz, CA 95064, USA}
\date{\today}

\begin{abstract}
Galaxy clusters identified in optical imaging surveys suffer from projection effects: Physically unassociated galaxies along a cluster's line of sight can be counted as its members and boost the observed richness (the number of cluster members).  To model the impact of projection on cluster cosmology analyses, we apply a halo occupation distribution model to $N$-body simulations to simulate the red galaxies contributing to cluster members, and we use the number of galaxies in a cylinder along the line of sight (counts in cylinders) to model the impact of projection on cluster richness.  We compare three projection models: uniform, quadratic, and Gaussian, and we convert between them by matching their effective cylinder volumes.  We validate our mock catalogs using SDSS \redmapper clusters' data vectors, including counts vs.~richness, stacked lensing signal, spectroscopic redshift distribution of member galaxies, and richness remeasured on a redshift grid.  We find the former two are insensitive to the projection model, while the latter two favor a quadratic projection model with a width of $\approx180 ~\hiMpc$ (equivalent to the volume of a uniform model with a width of 100 $\hiMpc$ and a Gaussian model with a width of 110 $\hiMpc$, or a Gaussian redshift error of $0.04$).  Our framework provides an efficient and flexible way to model optical cluster data vectors, paving the way for a simulation-based joint analysis for clusters, galaxies, and shear. 
\end{abstract}

\maketitle


\section{Introduction}

The abundance and spatial distribution of galaxy clusters are sensitive to the growth of structure and hold the promise of measuring $\sigma_8$ at different redshifts to subpercent level \cite{Weinberg13, Huterer15, Wu21}.  Galaxy clusters are most easily identified as high-density regions of red galaxies in the sky, and the number of galaxies associated with a cluster, called richness, can be used as a mass proxy \cite{Abell58}.  Wide-area optical imaging surveys, such as the Vera C.~Rubin Observatory Legacy Survey of Space and Time (LSST), \textit{Euclid}, and the Nancy Grace Roman Space Telescope's High Latitude Survey, are expected to identify more than 100,000 clusters.  In these imaging surveys, we do not have accurate distance measurements, and cluster finders such as \redmapper (red-sequence matched-filter probabilistic percolation) \cite{Rykoff14, Rykoff16} rely on the red sequence---a tight relation between color and magnitude for galaxies in clusters---to identify cluster member galaxies and estimate cluster redshifts.  Cluster members defined in this way inevitably suffer from projection effects; namely, physically unassociated galaxies along the line of sight that have color consistent with the red-sequence model can be counted as members \cite{Cohn07, Erickson11, NohCohn11, NohCohn12, Zu17, BuschWhite17, Wojtak18, SunayamaMore19}.

Projection effects are one of the key systematic uncertainties limiting current optical cluster cosmology studies \cite{DESY1CL, To21b, Park21, Sunayama23b}.  Projection effects change cluster richness and lensing coherently, leading to biased lensing signals for clusters selected by richness.  \citet{Wu22} use the mock \redmapper catalog derived from the {\sc Buzzard} simulations \cite{DeRose19Buzzard} to study the cluster lensing bias associated with richness selection, finding that a counts-in-cylinder richness with projection depth $\pm 30~\hiMpc$ mimics the selection bias of the \redmapper cluster finding algorithm.  One of the key uncertainties in that study is the underlying galaxy--halo connection; different galaxy models tend to lead to different amplitudes of selection bias.  
\citet{Zeng23} create mock catalogs based on a halo occupation distribution (HOD) model and counts-in-cylinder richness, using them to demonstrate that it is feasible to self-calibrate the lensing bias using the correlation functions between clusters, galaxies, and shear. 
\citet{Salcedo23} apply a HOD + counts-in-cylinder model to reanalyzed Dark Energy Survey Year 1 (DES-Y1) cluster lensing data, which was initially found to be in $5.6\sigma$ tension with \textit{Planck} \cite{DESY1CL}.  This study shows that the DES-Y1 cluster lensing is consistent with the \textit{Planck} cosmology, and previous studies may have suffered from inflexible modeling in richness.

\citet[][C19 hereafter]{Costanzi19projection} quantify the projection effects in the \redmapper clusters from the Sloan Digital Sky Survey (SDSS)  using richness remeasured on a redshift grid [$\lambda(z)$ hereafter], calculated as follows: after running the default \redmapper algorithm and determining the cluster redshift $\zcl$, one deliberately offsets the cluster redshift to $\zcl+\Dz$ and recalculates the richness using the red-sequence model at $\zcl+\Dz$.  If the member galaxies are spread across a wide redshift range, $\lambda(z)$ will be broad.  As we will verify in this paper, $\lambda(z)$ is proportional to the membership probability as a function of line-of-sight distance.

\citet[][M21 hereafter]{Myles21projection} quantify the projection effects of the same SDSS \redmapper clusters using a subsample of member galaxies that have spectroscopic redshift measurements.  They decompose the probability density function (PDF) of member galaxy redshifts into a cluster component and a projected component, fitting a double-Gaussian model to derive the parameters for each component.  They have found that the projected component decreases with increasing richness, and their results agree broadly with the {\sc Buzzard} simulation and \Costanzi.  This spectroscopic sample is introduced in \citet[][R15 hereafter]{Rozo15RM4} to study the membership probabilities of red spectroscopic galaxies and \redmapper members [also see \cite{Farahi16, Wetzell21}].

In this paper, we present a method for generating mock cluster catalogs with realistic projection effects and validate our method using the SDSS \redmapper clusters.  We aim to make computationally inexpensive mock catalogs with realistic projection effects to model the data vectors in cluster cosmology analyses.  We use an HOD model to simulate the red galaxies with color and magnitude consistent with that of \redmapper cluster members.  We then simulate the impact of projection on cluster richness by counting the galaxies in a cylinder along the line of sight (counts in cylinders).  We compare three different projection kernels (functions determining the membership probabilities vs.~the line-of-sight distances): uniform, quadratic, and Gaussian.  These kernels have been used in the literature in different contexts \cite{BuschWhite17, Sunayama20, Costanzi19projection, Myles21projection}, and we aim to compare them systematically with observations.  A key in this comparison is that the projection depths in different kernels need to be converted from one to another by matching their 3D effective cylinder volumes (Sec.~\ref{sec:pmem_models}).

We compare our simulated clusters with several observational results from the SDSS \redmapper clusters:  
(1) cluster abundance (number counts vs.~richness); 
(2) stacked cluster lensing; 
(3) the spectroscopic redshift distribution of member galaxies from \Myles;
(4) the richness remeasured on a grid of redshift, $\lambda(z)$, from \Costanzi.  
We will show that the first two are relatively insensitive to the projection model, while the other two are sensitive to the shape and depth of the projection kernel.

This paper is organized as follows.  
We describe our models and procedures for generating mock catalogs in Sec.~\ref{sec:mocks}.  
We then describe the SDSS \redmapper datasets in Sec.~\ref{sec:redmapper_catalog}.
In Sec.~\ref{sec:mock_vs_data}, we compare our model predictions with the SDSS results.
We discuss our results in Sec.~\ref{sec:discussion} and summarize in Sec.~\ref{sec:summary}.
In Appendix~\ref{app:color_dist}, we present the relation between cluster member colors and distance uncertainties.
In Appendix~\ref{app:alt_HOD}, we test an alternative HOD model.


\textit{Fiducial cosmology and units.} Throughout this work, we adopt the cosmology used in the Uchuu simulation \cite{Ishiyama21}: a flat $\Lambda$CDM based on the \textit{Planck} 2015 results \cite{Planck15cosmo}: 
$\OmegaM = 0.3089$,
$\OmegaB = 0.0486$,
$\sigma_8 = 0.8159$,
$\ns = 0.9667$, and 
$h = 0.6774$ (see the last column of their Table 4). 
We adopt the spherical overdensity halo mass definition $\Mtwo$, which encloses a mass density 200 times the mean matter density of the Universe.  Unless otherwise noted, all masses are in the unit of $\hiMsun$, and all distances are in the unit of comoving $\hiMpc$.

This paper uses line-of-sight comoving distances ($\Dchi$) to describe projection effects.  Previous studies have used a combination of redshift differences ($\Dz$), radial velocities ($v$), and line-of-sight distances, sometimes with an extra factor of $(1+z)$.  To facilitate comparison with previous studies, we convert both $\Dz$ and $v$ to $\Dchi$ via the nonrelativistic approximation
\beq
\frac{v}{c} (1+z) = 
\Dz
= \Dchi ~ \frac{E(z)}{3000~\hiMpc}  ~ ,
\eeq
where 
\beq
E(z) = \sqrt{\OmegaM(1+z)^3 + (1-\OmegaM)} ~ ,
\eeq   
for a flat universe.
For example, 
with our assumed cosmology, 
a redshift difference 
$\Dz = 0.03$ corresponds to 
$v \approx 8000~\kms$ and 
$\Dchi = 85~{\rm comoving}~\hiMpc$.

When converting $\Dz$ and $v$ to $\Dchi$, we include the small shifts due to the line-of-sight peculiar velocities of galaxies.  A typical cluster velocity dispersion of 1000 $\kms$ corresponds to $\Dchi \approx 10 ~\hiMpc$ or $\Dz\approx0.003$ at $z=0.1$, which is negligible compared with typical projection depths and photometric redshift uncertainties.

\section{Mock cluster catalogs}
\label{sec:mocks}

We aim to simulate not only galaxies physically associated with clusters (true members) but also galaxies located along the line of sight of clusters (projected members).  To this end, we populate halos of mass $\Mtwo > 10^{11}~\hiMsun$ with galaxies using HOD parameters motivated by luminous red galaxies.  We emphasize that the halo occupation is distinct from cluster richness; the halo occupation refers to galaxies within the 3D radius of halos, while the cluster richness includes projected members.

\subsection{The Uchuu \textit{\textbf{N}}-body simulation}

We use the publicly available $N$-body simulation Uchuu\footnote{\href{http://skiesanduniverses.org/Simulations/Uchuu/}{http://skiesanduniverses.org/Simulations/Uchuu/}} \cite{Ishiyama21}, which has a box size of $2~\hiGpc$, a mass resolution of $3.27\times10^8~\hiMsun$, and a gravitational softening length of $4.27~\hikpc$.  We also use the Mini Uchuu box, which has the same resolution with a smaller box size of $400~\hiMpc$.  The initial condition of the simulation is based on the {\sc 2lptic} code \cite{Crocce06}, the gravitational evolution is based on the {\sc greem} $N$-body code \cite{Ishiyama09}, and the halo finding is based on {\sc rockstar} \cite{Behroozi13rs}.

We use the $z=0.1$ output (snapshot 047) to simulate the SDSS \redmapper clusters in the redshift range $0.08<z<0.12$.   We use host halos with $\Mtwo \geq 10^{11}~\hiMsun$ for populating galaxies, and we use halos with $\Mtwo \geq 10^{12.5}~\hiMsun$ for mock cluster finding.   We subsample $5\times10^{-4}$ of the dark matter particles to calculate the gravitational lensing signal and to assign satellite galaxy positions and velocities.  We have tested that this subsampling gives consistent results as a $10^{-2}$ subsampling using Mini Uchuu.  To calculate error bars for the various predictions throughout the paper, we divide the Uchuu box into 250 subvolumes, each with a volume $0.4 \times 0.4 \times 0.2 = 0.032 ~(\hiGpc)^3$, similar to the volume of our SDSS sample.

\subsection{Assigning red galaxies to halos using an HOD model}

\begin{figure}
\centering
\includegraphics[width=\columnwidth]{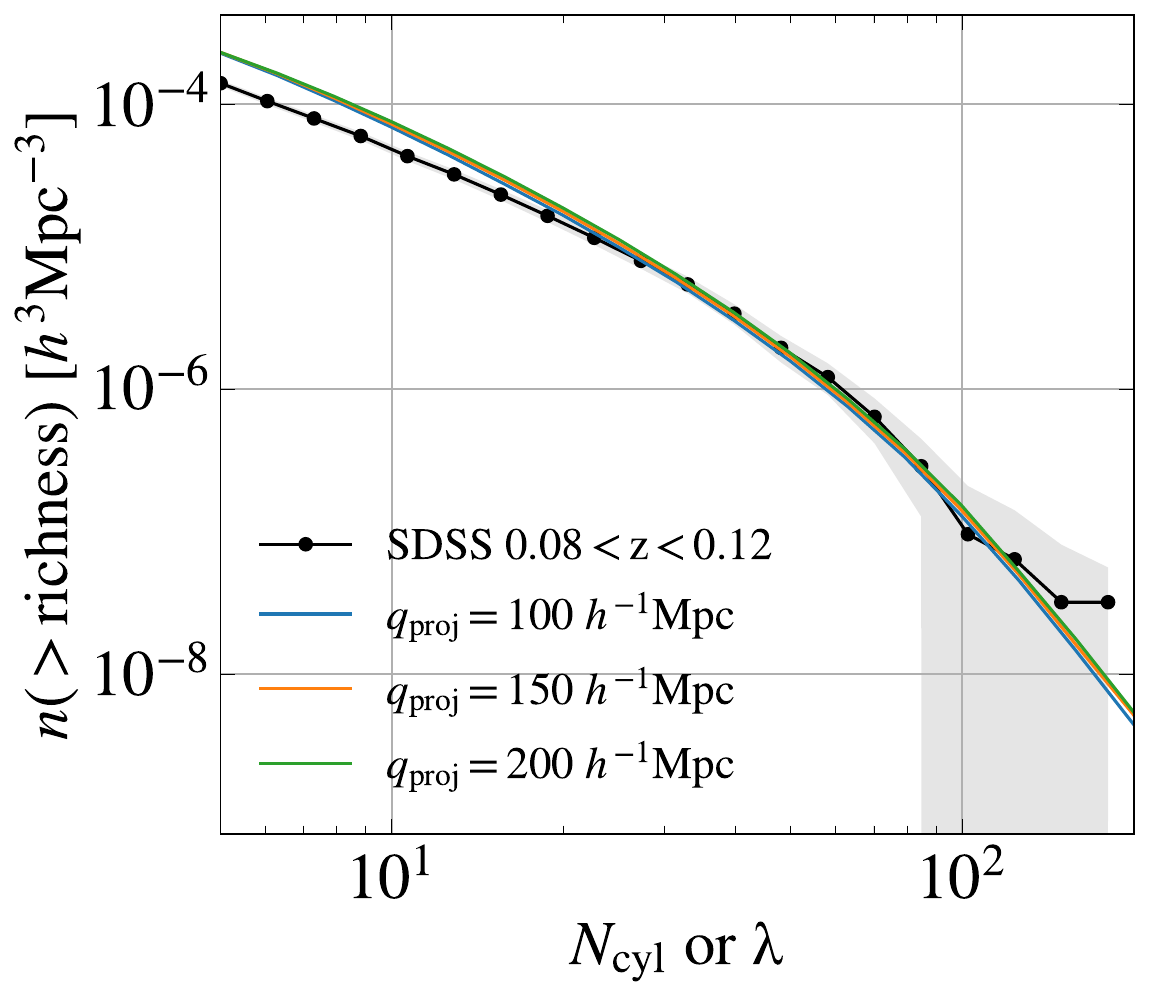}
\caption{Cumulative cluster number density as a function of richness.  The black curve shows SDSS \redmapper richness ($\lambda$), and the colored curves show the simulated counts-in-cylinders richness ($\Ncyl$).  We show the quadratic projection kernel [Eq.~(\ref{eq:pmem_quad})] with various depths, while other projection kernels behave similarly.  The number density is insensitive to the projection depth because of the background subtraction [Eq.~(\ref{eq:Ncyl})].  The gray band represents the statistical error bars estimated assuming the Poisson and halo sample variance.}
\label{fig:abundance}
\end{figure}

For a galaxy sample defined by a color-magnitude selection, the HOD specifies the number and spatial distribution of these galaxies in a given halo as a function of mass \cite{Seljak00, Benson00, PeacockSmith00, Scoccimarro01, GuzikSeljak02, BerlindWeinberg02, Berlind03, vdBosch03, Kravtsov04, Yang04, Zheng05}.  We use an HOD model to simulate red-sequence galaxies with colors and magnitudes consistent with \redmapper members.  For a given halo mass, we assume that the mean number of central galaxies is given by
\beq
\langle \Ncen | M\rangle 
= \frac{1}{2} 
\left[
1 + {\rm erf}\left(
\frac{\log_{10} M - \log_{10} \Mmin}{\sigma_{\rm \log M}}
\right)
\right] ~,
\eeq
and the mean number of satellite galaxies is given by
\beq
\langle \Nsat | M \rangle  = 
\left\{ 
 \begin{array}{cl}
\left( \frac{M - M_0}{M_1} \right)^{\alpha}~, & \mathrm{if~} \Ncen = 1~, \\
 0~, & \mathrm{otherwise}~.
 \end{array} 
 \right.
\eeq
We adopt the parameters in \citet{Sunayama20} as our fiducial model:  $\Mmin = M_0 = 10^{11.7} \hiMsun$,  $M_1 = 10^{12.9} \hiMsun$, and $\alpha=1$.  This model reproduces the SDSS \redmapper cluster number counts vs.~mass (see Fig.~\ref{fig:abundance}). 

We first determine whether each halo has a central galaxy by drawing an $\Ncen$ value (0 or 1) based on a Bernoulli distribution with probability $\avg{\Ncen|M}$.  If a halo has $\Ncen=1$, we place a central galaxy at the halo center and assign it the velocity of the halo.  We determine the number of satellite galaxies in this halo by drawing an $\Nsat$ value based on a Poisson distribution with mean $\avg{\Nsat|M}$.  We then randomly draw $\Nsat$ dark matter particles within $\Rtwo$ and assign their positions and velocities to these satellite galaxies.

Because of the $5\times10^{-4}$ subsampling of dark matter particles, sometimes a low-mass halo has no particles.  These halos usually have zero satellites, but in some rare cases, they can have one or two satellites.  In this case, we draw the satellite positions using an NFW profile \cite{NFW97} and draw the velocities using a Gaussian distribution with the velocity dispersion from \citet{Evrard08} 
\beq
\log\sigma
= \log(1082.9 ~{\kms})
+ 0.3361 \log
\left(\frac{E(z)M_{\rm 200c}}{10^{15}\hiMsun} \right) ~,
\eeq
converting from $\Mtwo$ to $M_{\rm 200c}$ assuming an NFW profile and the concentration-mass relation from \citet{Bhattacharya13}.  We have tested that using the analytic formulas for positions and peculiar velocities gives similar results to drawing dark matter particles.  We have also compared this approach with using 1\% particles in Mini Uchuu and found negligible differences.

\subsection{Simulating cluster finding}

In this step, we simulate the projection effects in the cluster finding process by applying counts in cylinders to the mock galaxy catalogs \cite{Angulo12, BuschWhite17, Sunayama20}.  We use the centers of halos with $\Mtwo \ge 10^{12.5}\hiMsun$ as our cluster centers and defer the modeling of miscentering \cite{Zhang19, Kelly23} to future work.  We search for galaxies near the line of sight of these cluster centers.  We assume that a galaxy within the projected radius $\Rcyl$ has a membership probability $\pproj(\Dchi)$, where $\Dchi$ is the line-of-sight distance between the galaxy and the cluster center.  We refer to $\pproj(\Dchi)$ as the projection kernel and will present several models in Sec.~\ref{sec:pmem_models}.  We assume no radial dependence of $\pproj(\Dchi)$ inside $\Rcyl$ and exclude all galaxies outside $\Rcyl$.

The richness of a cluster is the sum of the $\pproj(\Dchi)$ of all galaxies with a projected distance $\rp \leq \Rcyl$, with the mean number of background galaxies subtracted,
\beq
\Ncyl = \sum_{\rp < \Rcyl} \pproj(\Dchi)  - N_{\rm bg}~,
\label{eq:Ncyl}
\eeq
where $N_{\rm bg}$ is the product of the mean galaxy number density and the effective cylinder volume associated with the projection kernel (defined in Sec.~\ref{sec:pmem_models}).

Following the definition of \redmapper's cluster radius $\Rlam$, we define the projected cluster radius as
\beq
\Rcyl =  \left(\frac{\Ncyl}{100}\right)^{0.2}
~ \mbox{physical} ~\hiMpc ~,
\label{eq:Rcyl}
\eeq
using physical $\hiMpc$ to match the convention of \redmapper.  This radius-richness scaling has been chosen to minimize the scatter of richness at a given x-ray luminosity \cite{Rykoff12}.  In the future, we will revisit the optimal choice of the projected cluster radius.

For a given cluster center, we solve $\Ncyl$ and $\Rcyl$ in Eq.~(\ref{eq:Ncyl}) and Eq.~(\ref{eq:Rcyl}) iteratively.  When a galaxy is within the cylinders of multiple clusters, its contribution to a higher-ranking cluster is prioritized.  We mimic the probabilistic percolation process of \redmapper by ranking our cluster centers by their halo masses.  When a galaxy is assigned as a member of a cluster with membership probability $\pproj$, this galaxy's availability to be a member of a lower-ranked cluster is reduced by $\pproj$; when its availability is below 20\%, this galaxy is removed from the pool of available member galaxies.

\subsection{Modeling projection effects}
\label{sec:pmem_models}

We consider three models for the projection kernel $\pproj(\Dchi)$, the membership probability as a function of line-of-sight distances.  In each model, we define the effective cylinder volume as 
\beq
V_{\rm cyl} = 
\int_{-\infty}^{\infty} 2\pi \Rcyl^2 ~ \pproj(x)^2 dx ~.
\label{eq:Vcyl}
\eeq
{\noindent \it Model I. Uniform.} 
This model counts all galaxies inside a cylinder within a projected radius $\Rcyl$ and a line-of-sight comoving distance $\pm~\dproj~\hiMpc$ with equal weights.
\beq
 \pproj(\Dchi) = 
 \left\{ 
 \begin{array}{cl}
 1 ~ , & \mathrm{if}~|\Dchi|\leq \dproj ~,\\
 0 ~ , & \mathrm{otherwise}~.
 \end{array} 
 \right.
\label{eq:pmem_uniform}
\eeq
The volume of the uniform projection kernel is $2 \pi \Rcyl^2 \dproj$.

\bigskip

{\noindent \it Model II. Quadratic.} 
Motivated by \Costanzi, we use a quadratic function to weight galaxies according to their line-of-sight distances. 
\beq
 \pproj(\Dchi) = 
 \left\{ 
 \begin{array}{cl}
 1-\left(\frac{\Dchi}{\qproj}\right)^2 ~ , & \mathrm{if~} \Dchi\leq \qproj  ~ , \\
 0 ~ , & \mathrm{otherwise}~.   
 \end{array} 
 \right. 
\label{eq:pmem_quad}
\eeq
Here $\pproj(\Dchi=0)=1$. The effective cylinder volume of the quadratic projection kernel is $(16/15) \pi  \Rcyl^2 \qproj$.

\bigskip

{\noindent \it Model III. Gaussian.} 
Motivated by the double Gaussian distribution of member galaxies' spectroscopic redshifts found in \Myles, we assume a Gaussian-shaped function with a width $\sigproj$,
\beq
\pproj(\Dchi) = e^{-\frac{1}{2}\left(\frac{\Dchi}{\sigproj}\right)^2} ~,
\eeq
so that $\pproj(\Dchi=0)=1$. The volume of the Gaussian projection kernel is $\pi^{3/2} \Rcyl^2 \sigproj$.

\bigskip

Each model has a characteristic projection depth: $\dproj$, $\qproj$, or $\sigproj$.  To compare these models, we match their effective cylinder volumes  [Eq.~(\ref{eq:Vcyl})]. We reason that if galaxies are uniformly distributed, projection kernels with the same volume should include the same number of galaxies.  Under this volume matching,  $\sigproj=90~\hiMpc$ is equivalent to $\dproj=80~\hiMpc$ and $\qproj=150~\hiMpc$.  As a rule of thumb, $\dproj$ and $\sigproj$ have similar values, while $\qproj$ is approximately twice as large as $\dproj$.

\begin{figure}
\includegraphics[width=\columnwidth]{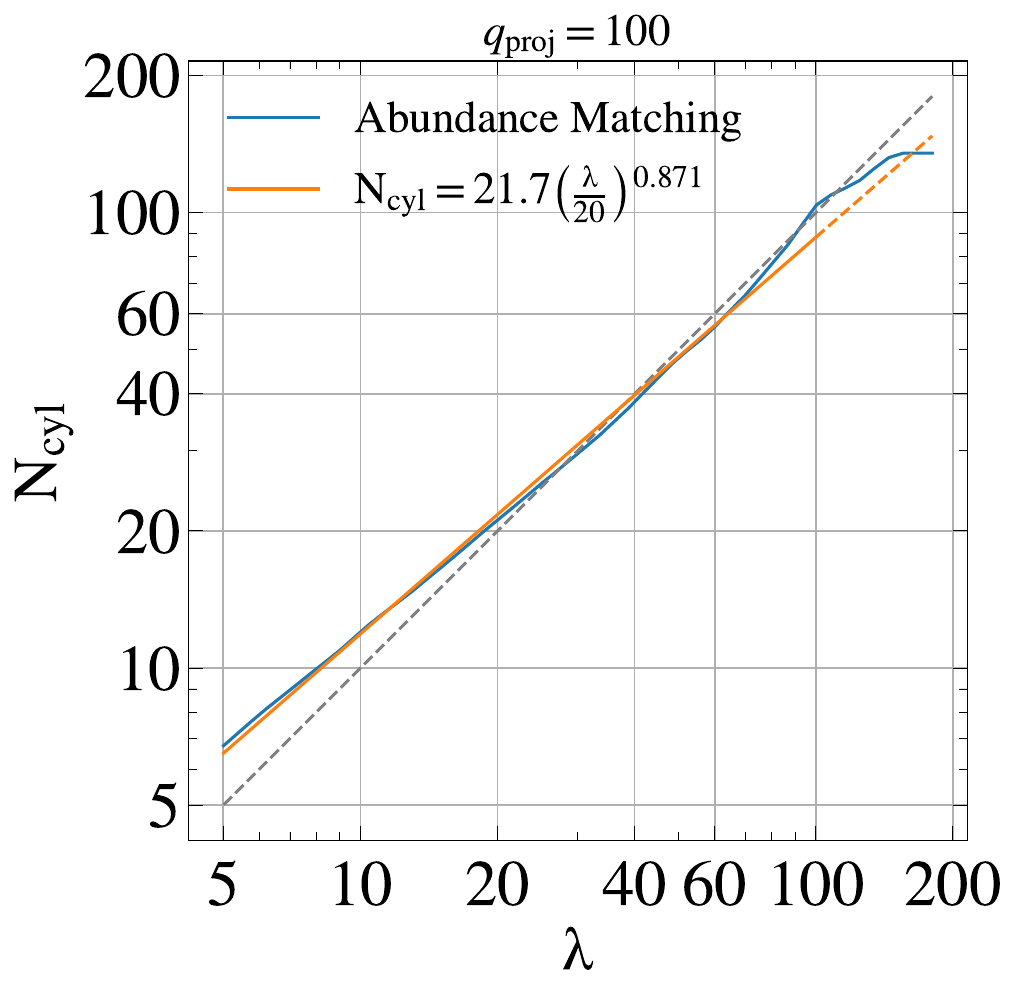}
\caption{An example of the relation between the SDSS \redmapper richness $\lambda$ and the simulated counts-in-cylinder richness $\Ncyl$ (blue curve) derived from abundance matching [Eq.~(\ref{eq:abun_match})].  We show a quadratic projection kernel with $\qproj=100~\hiMpc$.  The gray dashed line shows the 45$^\circ$ line.  As seen in Fig.~\ref{fig:abundance}, $\Ncyl$ closely resembles $\lambda$ for $\lambda\gtrsim20$ and is lower for $\lambda\lesssim20$.  The orange line shows the best-fit power law for $5 \leq \lambda \leq 100$.  This power law is used in the high-richness regime, where abundance matching is noisy.}
\label{fig:lambda_Ncyl}
\end{figure}

Figure~\ref{fig:abundance} shows the cumulative number density of the simulated clusters as a function of $\Ncyl$.  We show the quadratic model with projection depths $\qproj$ = 100, 150, and 200 $\hiMpc$; the uniform and Gaussian models behave similarly.  With the background subtraction, $\Ncyl$ is insensitive to the projection depth between 100 and 200 $\hiMpc$; this suggests that most correlated structures contributing to cluster richness fall within 100 $\hiMpc$.  A much smaller projection depth (e.g., $\qproj=10~\hiMpc$) will underpredict the richness.  For $\Ncyl \gtrsim 20$, our simulations predict a cluster abundance similar to SDSS; for $\Ncyl \lesssim 20$, our simulated richness tends to be higher than SDSS richness, which leads to a higher cumulative abundance.  To account for the small discrepancy between $\lambda$ and $\Ncyl$, we will perform abundance matching when comparing them (Sec.~\ref{sec:mock_vs_data} and Fig.~\ref{fig:lambda_Ncyl}).  
The gray band corresponds to the statistical error due to Poisson shot noise and halo sample variance,  analytically calculated based on Ref.~\cite{HuKravtsov03}.

\section{SDSS \redmapper cluster catalog}
\label{sec:redmapper_catalog}

To validate our mock catalogs, we use clusters in the SDSS DR8 \redmapper v5.10 catalog introduced in \Rozo.  We briefly review the \redmapper cluster finding algorithm and refer readers to Refs.~\cite{Rykoff14, Rykoff16} for details.  The \redmapper cluster finder identifies galaxy clusters based on overdensities of red-sequence galaxies.  The algorithm includes two stages: the red-sequence calibration and the cluster finding.  In the first stage, \redmapper calibrates the red-sequence model, i.e., the color-magnitude relation of cluster galaxies as a function of redshift, using cluster samples with known redshifts.  In the second stage, \redmapper searches for overdensities of galaxies with colors consistent with the red-sequence model at a given redshift.  These galaxies are assigned a membership probability based on their color, magnitude, and projected distance to the cluster center.  The richness $\lambda$ is defined as the sum of the membership probability of galaxies within an aperture.  After the initial red-sequence calibration stage, \redmapper iteratively calibrates the red-sequence models and finds clusters.  The cluster redshift $z_\lambda$ is iteratively determined by maximizing the member galaxies' color likelihood.

Searching for galaxies with colors consistent with the red-sequence model at a given redshift is similar to estimating photometric redshifts.  The redshift uncertainties of cluster members are determined by the width of the red sequence and by how fast the red sequence changes with redshift.  These redshift uncertainties are the origin of the projection effects discussed in this work.

In \Rozo, the SDSS DR9 \redmapper cluster galaxies are matched to the spectroscopic samples from SDSS DR10 \cite{SDSS_DR10} and GAMA \cite{Driver09} based on an angular matching of 1 arcsecond.  Following \Myles, we select from the \redmapper SDSS catalog clusters with  $\lambda > 5$ and $0.08 \leq z \leq 0.12$, covering an area of 10,400 deg$^2$.  For our fiducial cosmology, the comoving volume of this dataset is 0.032 $(\hiGpc)^3$.

When using the spectroscopic data, we select clusters with available central galaxy spectroscopic redshifts $\zCG$, and we use $|\zCG - \zlam| \leq 0.03$ to remove misidentified central galaxies.  The resulting cluster catalog has 2439 clusters.  We select member galaxies with $L \geq 0.55 L_*$ in the $i$ band, while the full \redmapper members have $L \geq 0.2 L_*$.  According to \Myles, this $0.55 L_*$ cut corresponds to $\approx 37\%$ of the total members, and the spectroscopic completeness of this sample is $\approx 84\%$.


\section{Comparing our models with SDSS \redmapper clusters} \label{sec:mock_vs_data}

This section compares our model prediction with three data vectors derived from the SDSS \redmapper clusters: stacked cluster lensing, spectroscopic redshift distribution of member galaxies, and $\lambda(z)$.

Given that $\Ncyl$ is a simplified version of $\lambda$, we do not expect their values to agree exactly (see Fig.~\ref{fig:abundance}); instead, we compare mock clusters ranked by $\Ncyl$ with observed clusters ranked by $\lambda$.  This is the essence of abundance matching: We associate the $\Ncyl$ threshold with the $\lambda$ threshold that gives the same number density.  We first calculate the cumulative number densities as a function of $\lambda$ or $\Ncyl$ threshold.  For a given $\lambda$, we find $\Ncyl$ that satisfies the relation
\beq
n(> \lambda) = n(> \Ncyl) ~.
\label{eq:abun_match}
\eeq
Figure~\ref{fig:lambda_Ncyl} shows the $\lambda$--$\Ncyl$ relation derived from this abundance matching procedure for a quadratic projection kernel with $\qproj$ = 100 $\hiMpc$.  We show the best-fit power law in the legend.  For $\lambda > 100$, where the number density is low and the abundance matching is noisy, we use this power law to assign $\lambda$ to each simulated cluster to reduce noise.  We repeat this procedure for different projection models.  Unless otherwise noted, all the simulated richness in this paper refers to the abundance-matched one.

\subsection{Stacked weak lensing}
\begin{figure}
\includegraphics[width=\columnwidth]{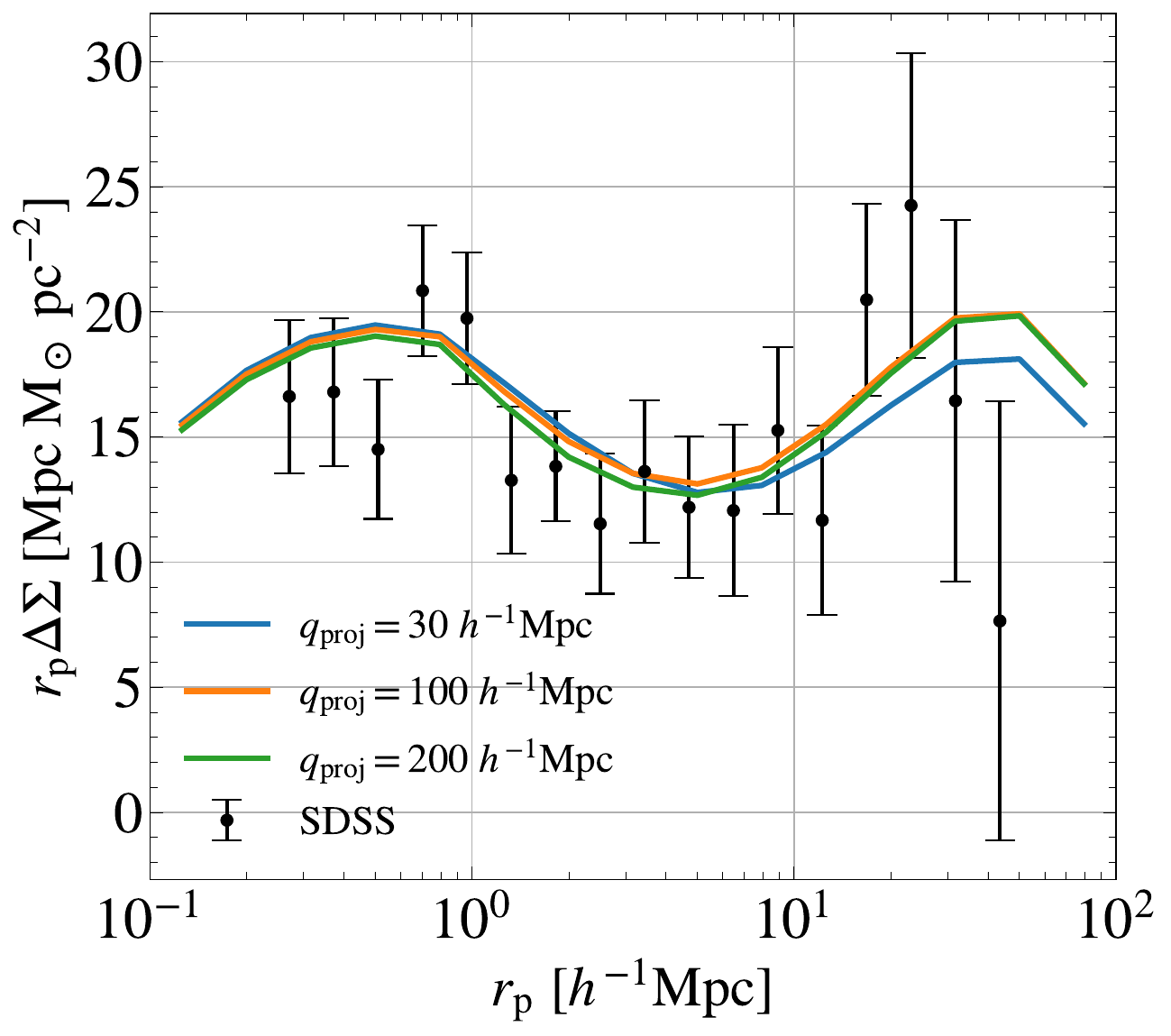}
\caption{Weak lensing signal of SDSS clusters with $\lambda > 5$ and $0.08 < z < 0.12$ (black points with error bars) compared with the predictions from our mock catalogs with matched abundance (colored curves).  We show the quadratic projection kernel with $\qproj$ = 30, 100, and 200 $\hiMpc$, while all other kernels behave similarly.  The lensing signal is insensitive to the projection depth.}
\label{fig:lensing}
\end{figure}
\begin{figure*}
\centering
\includegraphics[width=2\columnwidth]{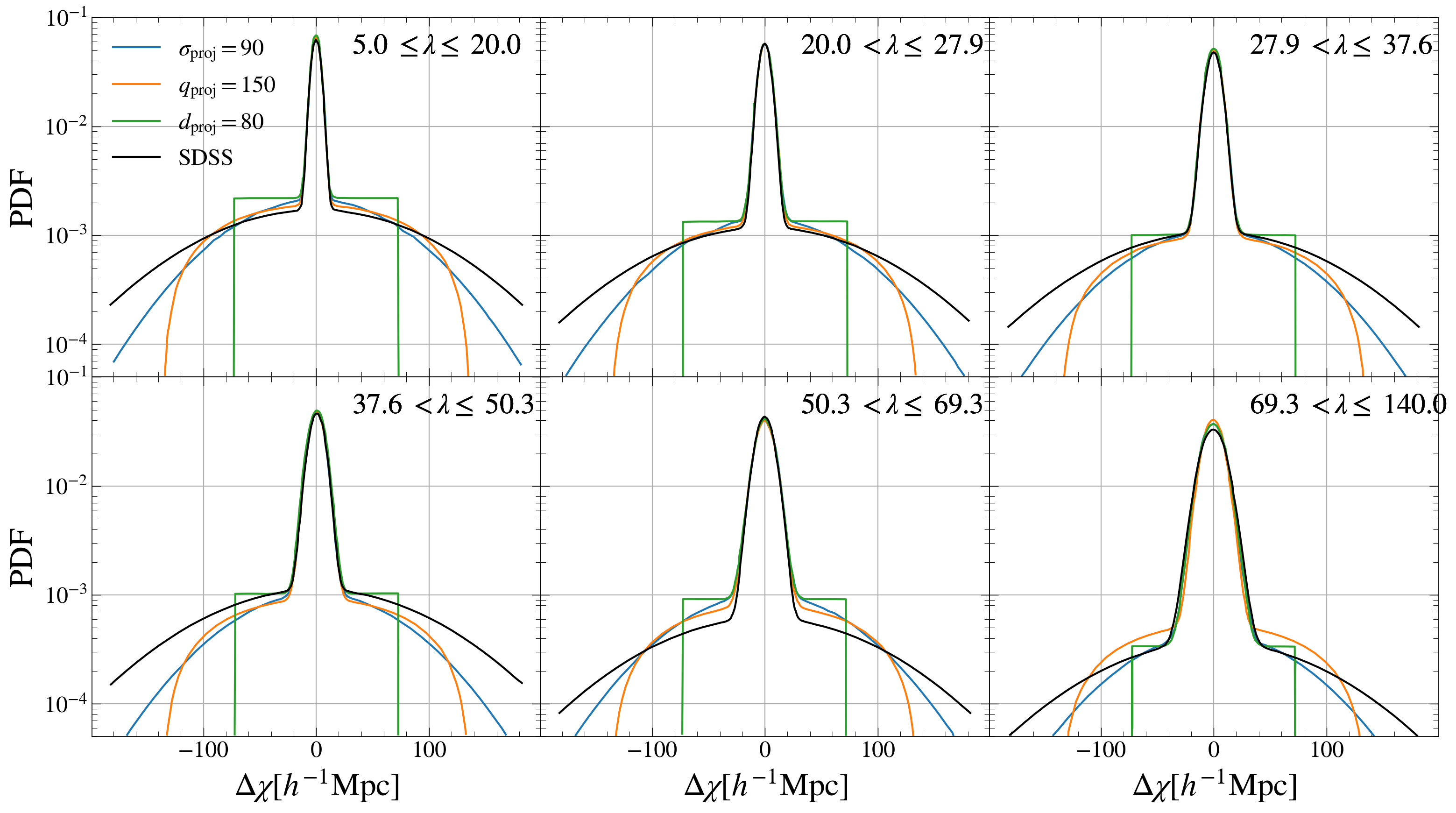}
\caption{Best-fit two-component PDF for the spectroscopic redshift distribution of cluster members (data not shown to avoid crowding), in six abundance-matched richness bins.  The black curves show the best-fit PDF for SDSS (Fig.~2 in \protect\Myles), and the colored curves show the best-fit PDF for our mock catalogs [Eq.~(\ref{eq:P_Dchi})]; note that the colored curves are fit to the mock data rather than to the black curves.  We show the uniform (green), quadratic (orange), and Gaussian (blue) projection kernels with matched effective cylinder volumes (Sec.~\ref{sec:pmem_models}).  The uniform model deviates most from the SDSS data,
and we will focus on quadratic and Gaussian models for the rest of this paper.
}
\label{fig:delz_hist}
\end{figure*}

We measure the stacked weak lensing signal of the SDSS DR8 \redmapper v5.10 sample, with $\lambda > 5$ and $0.08 < z_\lambda < 0.12$, covering 10,400 $\mathrm{deg}^2$ and including 4,390 clusters.   We do not require spectroscopic redshifts when selecting these clusters.  Assuming $\OmegaM=0.3089$, the comoving number density of this sample is $1.4\times10^{-4} ~\densityunit$.  To calculate the lensing signal from SDSS, we follow the same procedure as in \citet{Park21}.  Below, we briefly summarize the measurement procedure.  We use the estimator 
\beq
\Delta\Sigma(\rp)
= \frac{\sum_l\sum_s w_{ls} e_{{\rm t},ls} \Sigma_{{\rm crit},ls}}{\big(1+\hat{m}(\rp)\big)\sum_l\sum_s w_{ls}} \ ,
\eeq
where $\rp$ is the projected distance to the cluster center, $l$ and $s$ indicate lens and source bins, $e_{{\rm t},ls}$ is the tangential ellipticity for a given lens-source pair, $\Sigma_{{\rm crit},ls}$ is the critical surface density for this lens-source pair, $\hat{m}$ is the multiplicative shear bias, and $w_{ls}$ indicates the lens-source pair weights 
\beq
w_{ls} = w_l  w_s \Sigma_{{\rm crit},ls}^{-2}  ~.
\eeq
The error bars are calculated using 83 independent Jackknife regions.

To calculate the lensing signal from our mock catalogs, we choose clusters above a richness threshold to match SDSS \redmapper clusters' comoving number density at $\lambda > 5$.  We calculate $\Delta\Sigma$ by calculating the cross-correlations between clusters and dark matter particles using the {\sc Corrfunc} software \cite{Corrfunc}.

Figure~\ref{fig:lensing} compares the SDSS \redmapper cluster weak lensing signal (black points with error bars) with our model predictions.  We show the quadratic kernel with depths $\qproj$ between 30 and 200 $\hiMpc$; other projection kernels behave similarly.  Our predicted lensing signal agrees with the observed one and is insensitive to the projection depth.  The 30 $\hiMpc$ projection kernel predicts a lower lensing signal at large scales, which is still consistent with the SDSS error bars.  The SDSS error bars are dominated by shape noise; since our simulations have no shape noise, we do not show their error bars.  Because we use a low richness threshold $\lambda > 5$ and match the abundance, these mock catalogs include similar halo populations and lensing signals.  When using several richness bins, our predictions are still consistent with the observations.

The SDSS lensing has a rather large noise (with a source galaxy density of 1.2 arcmin$^{-2}$ \cite{Reyes12}) and cannot distinguish the models we test here. In addition, the relatively low redshifts of SDSS sources may have led to unknown systematics; for example, \citet{Sunayama23b} have shown that the cluster lensing based on SDSS sources is systematically lower than that based on Hyper-Suprime Cam (HSC) sources (see their Appendix C).  We expect that high-resolution lensing measurements from deep surveys such as HSC (with a source density of 20 arcmin$^{-2}$ \cite{Li22HSC}), as well as the upcoming LSST and Roman data, would be able to distinguish different projection models.

\subsection{Spectroscopic redshift distribution of member galaxies}
\label{sec:Pz}

\begin{figure*}
\includegraphics[width=\columnwidth]{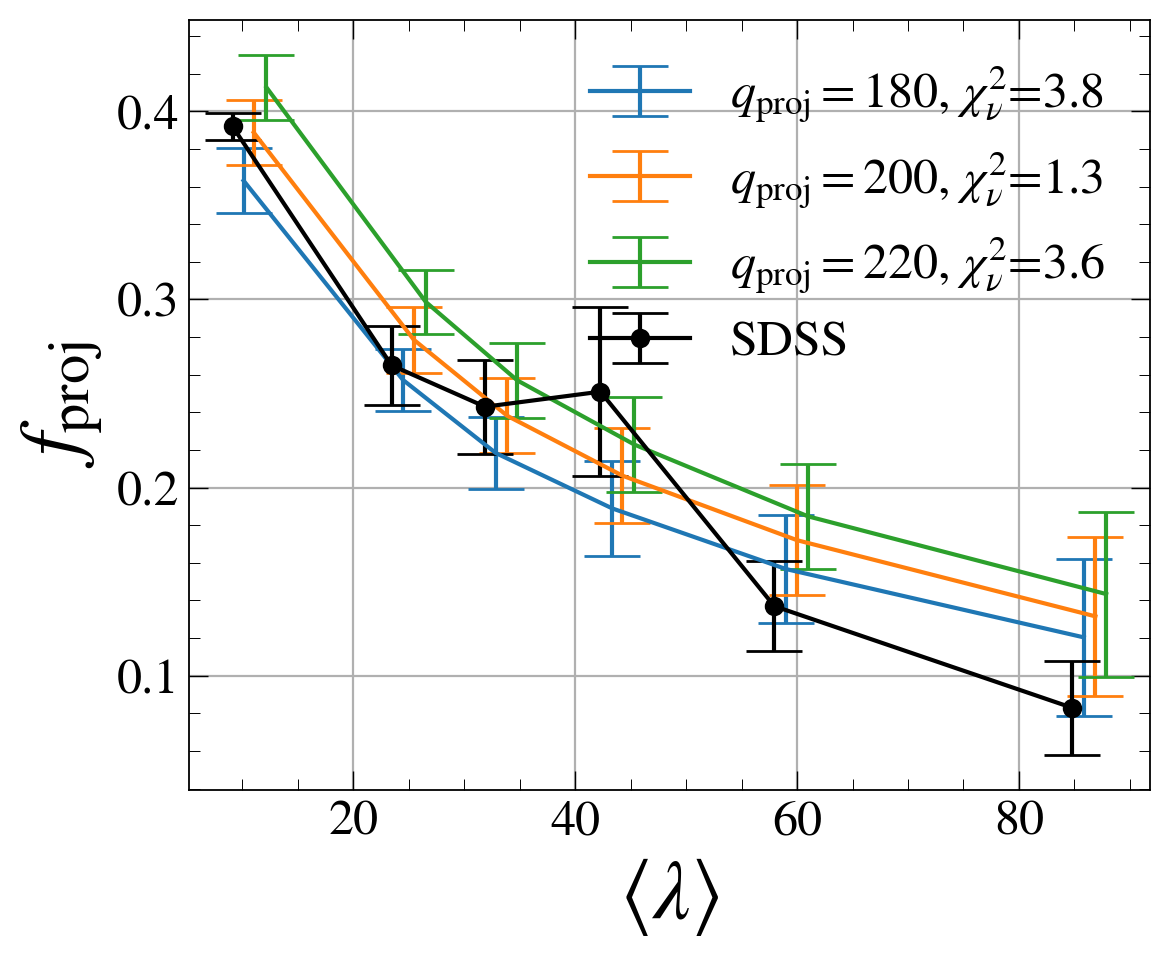}
\includegraphics[width=\columnwidth]{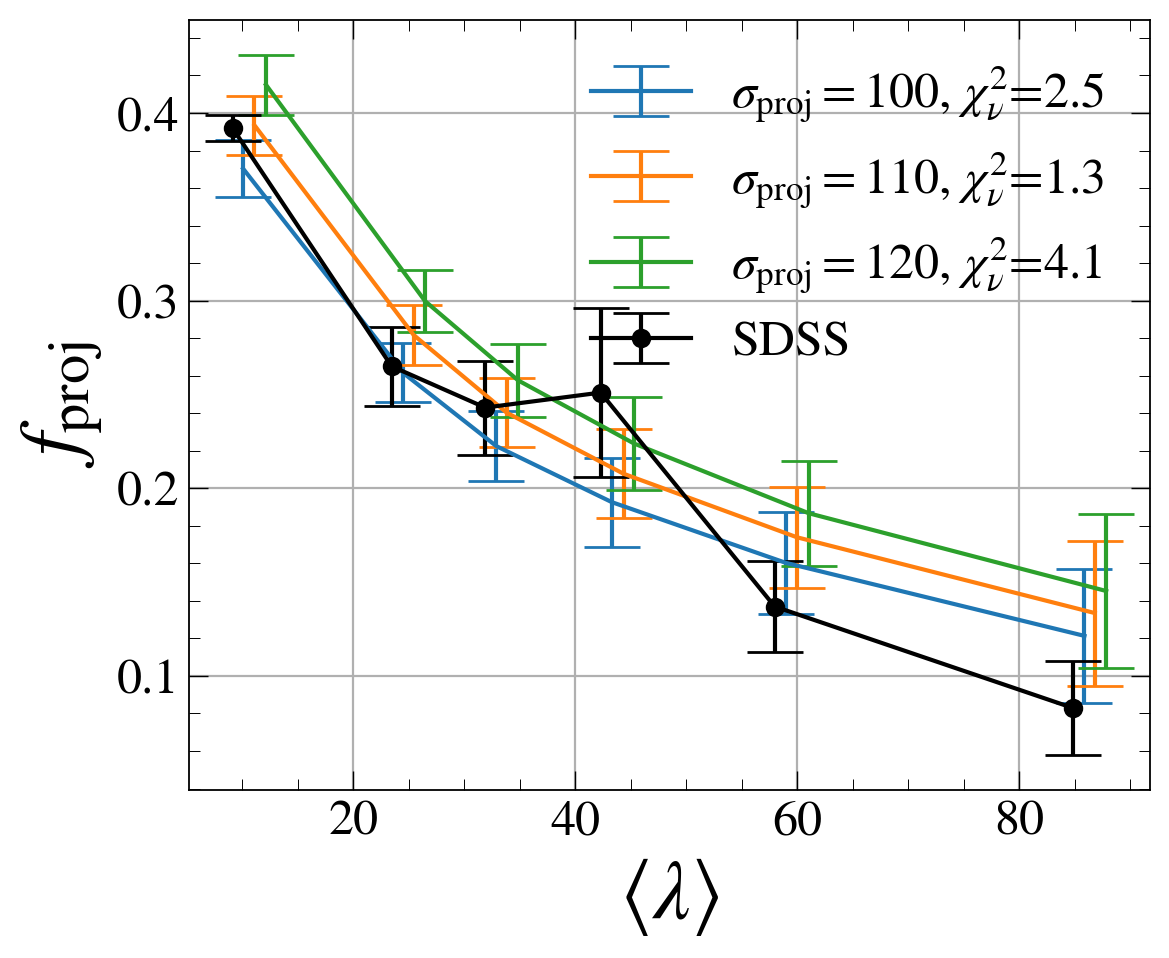} 
\caption{Projection fraction $\fproj$ derived from the best-fit two-component PDF as a function of richness.  The black curves show the results from SDSS (\protect\Myles), and the colored curves show the results from our mock catalogs.  The left-hand panel shows the quadratic model, and the right-hand panel shows the Gaussian model.  Our mock catalogs capture the decreasing $\fproj$ with richness, and the amplitude of $\fproj$ is sensitive to the projection depth.  The SDSS data is consistent with $\qproj$ = 180 $\hiMpc$ and $\sigproj$ = 100 $\hiMpc$, which have approximately the same effective cylinder volumes. }
\label{fig:fproj}
\end{figure*}
\begin{figure}
\centering
\includegraphics[width=\columnwidth]{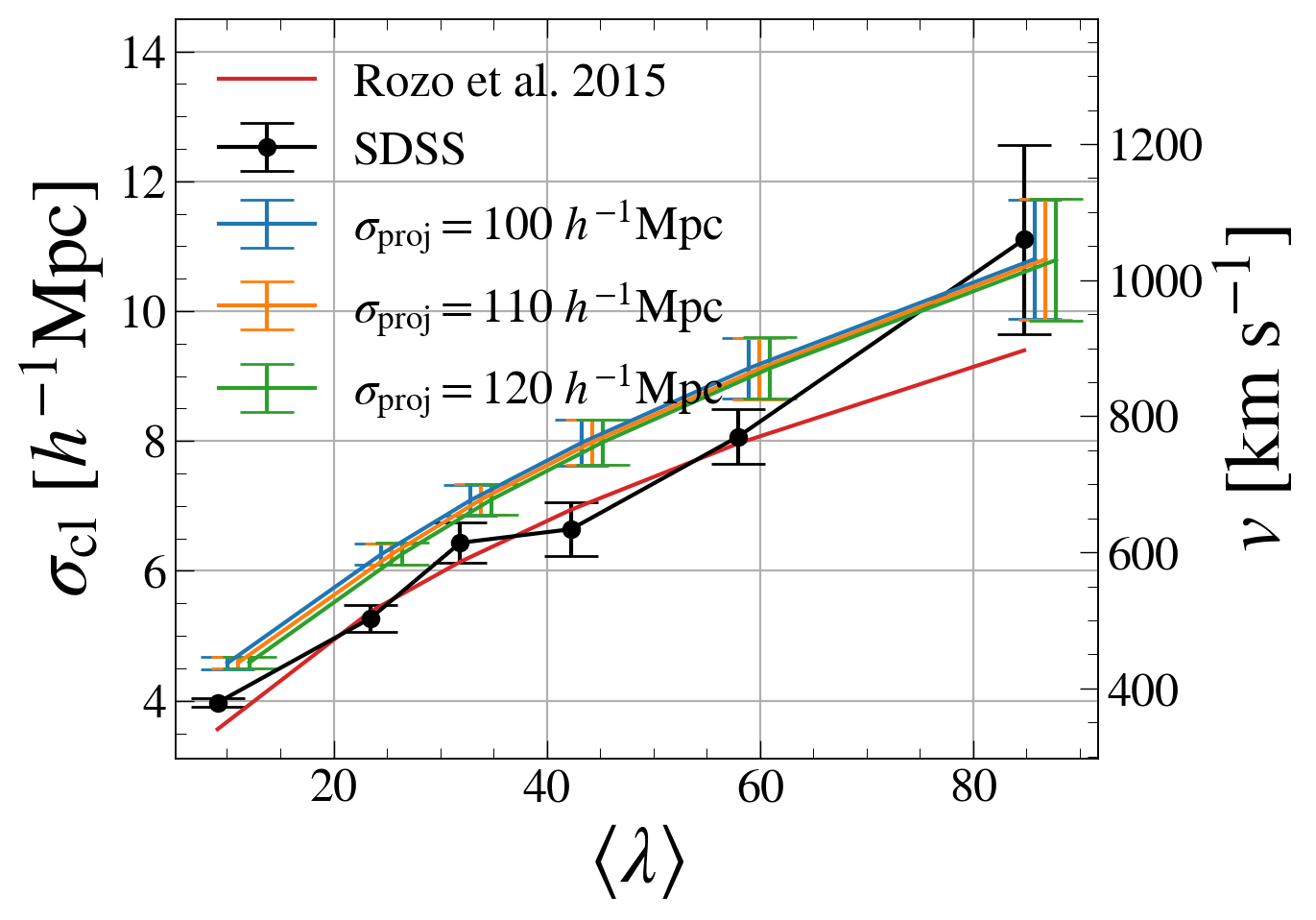}
\caption{The central Gaussian component width $\sigcl$ (the velocity dispersion of galaxies in clusters) derived from the best-fit two-component PDF.  The right $y$ axis shows line-of-sight velocity, while the left $y$ axis shows the corresponding line-of-sight comoving distance.  The black curve shows the SDSS result, and the colored curves show our mocks (the same models as in the right-hand panel of  Fig.~\ref{fig:fproj}).  The red curve shows the velocity dispersions of SDSS \redmapper clusters derived in \protect\Rozo.  Our mock catalogs slightly overpredict the velocity dispersion, which is insensitive to the projection model.}
\label{fig:sigma_cl}
\end{figure}

\begin{figure*}
\centering
\includegraphics[width=2\columnwidth]{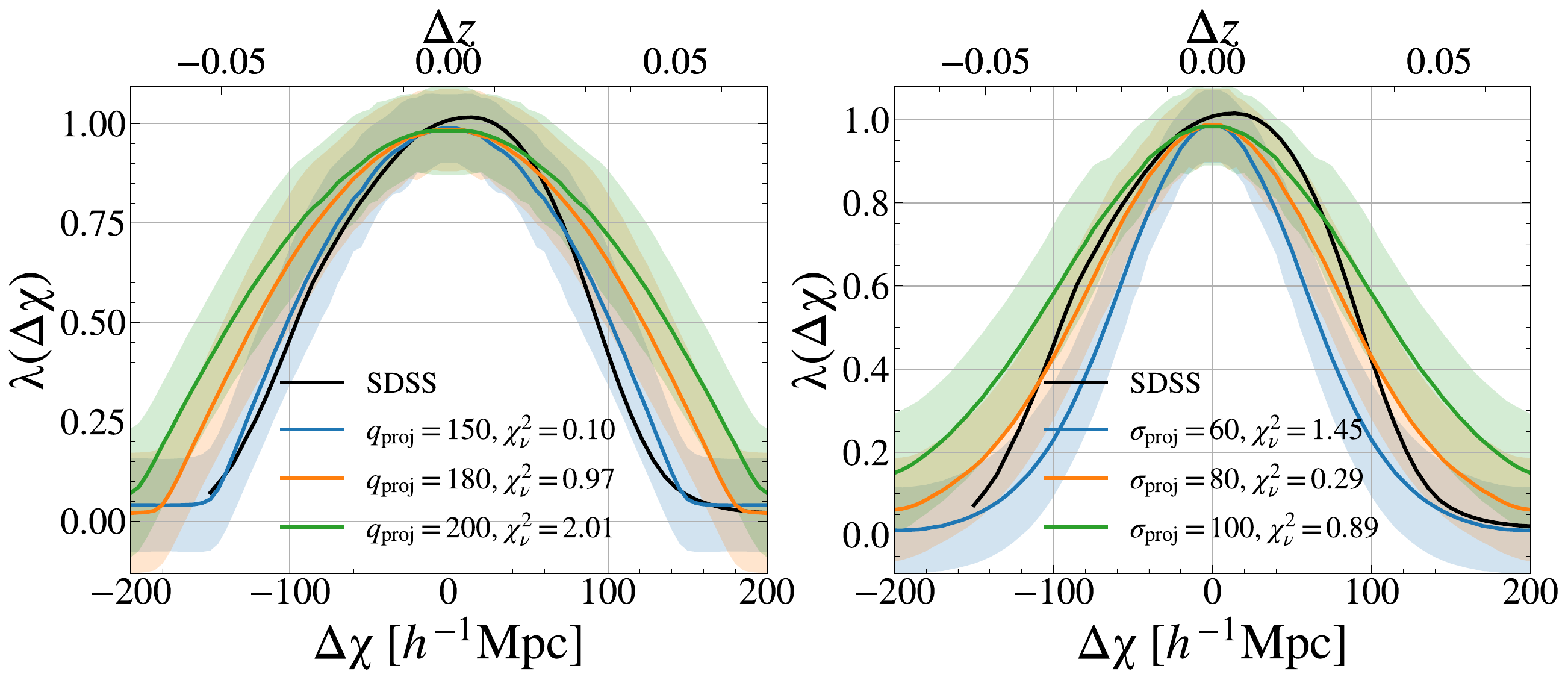}
\caption{The richness recalculated on a redshift grid $\lambda(z)$, plotted as a function of redshift difference (top $x$ axes) and line-of-sight comoving distance (bottom $x$ axes).  The black curves show the measurements from SDSS, and the colored curves correspond to our mock catalogs.  The left-hand panel corresponds to the quadratic model, while the right-hand panel corresponds to the Gaussian model.  The quadratic model with $\qproj=150~\hiMpc$ agrees well with the observation, while the Gaussian model tends to be too wide in the outskirts.}
\label{fig:lambda_z}
\end{figure*}

In \Myles, the authors have calculated the spectroscopic redshift distribution of \redmapper members and fit it with a double-Gaussian model.  To compare our mocks with their results, we find the member galaxies of our mock clusters and plot the PDF of their line-of-sight distances from the cluster center.  To fit this PDF, we use several two-component models, including Gaussian+Gaussian (as in \Myles), Gaussian+uniform, and Gaussian+quadratic.  We detail our procedures below.

For a given projection kernel and depth parameter described in Sec.~\ref{sec:pmem_models}, we generate a mock cluster catalog and perform abundance matching between $\lambda$ and $\Ncyl$.  We collect all the members for clusters in each richness bin and plot the PDF of their line-of-sight distances, denoted as $\mathcal{P}(\Dchi)$.  In this calculation, each member is weighted by its membership probability $\pproj$.  We fit the PDF with a two-component model: 
\beq
\mathcal{P}\big(\Dchi\big) = 
\fcl ~ \mathcal{N}\big(\Dchi \big| 0, \sigcl\big)
+ \fproj ~ \mathcal{P}_{\rm proj}\big(\Dchi \big| {\rm depth}\big)~,
\label{eq:P_Dchi}
\eeq
where $\fcl + \fproj = 1$.  The first term corresponds to the cluster component, assumed to be a Gaussian distribution with a mean of zero and a standard deviation of $\sigcl$.  The second term corresponds to the projected component and is based on the normalized PDF of one of the three projection kernels described in Sec.~\ref{sec:pmem_models}.  For example, we use a normalized quadratic PDF for $\mathcal{P}_{\rm proj}$ for a mock generated with the quadratic projection kernel, and so on.  We jointly fit the PDF $\mathcal{P}(\Dchi)$ in six richness bins, using one $\fproj$ parameter and one $\sigcl$ parameter for each bin but the same depth parameter for all bins, following \Myles.  We have tested that using a different projection depth for each richness bin gives similar results.

Figure~\ref{fig:delz_hist} shows the normalized PDF of the member galaxies' line-of-sight distances for six abundance-matched richness bins.  The black curve shows the \Myles results, while the colored curves correspond to different $\pproj$ models presented in Sec.~\ref{sec:pmem_models}.  These colored curves are the best-fit two-component functions of each mock's histogram; they are not fit to the \Myles data.  Although all of these models deviate from the SDSS data at the tails of the PDF, the tails contribute to a negligible fraction of the projected population.  Given that the uniform model (green) deviates the most from the observation, we will focus on the quadratic and Gaussian models below.

The left-hand panel of Fig.~\ref{fig:fproj} shows the projected fraction $\fproj$ from the quadratic model with $\qproj=$ 180, 200, 220 $\hiMpc$, while the right-hand panel shows $\fproj$ from the Gaussian model with $\sigproj = $ 100, 110, 120 $\hiMpc$.  We obtain $\fproj$ by minimizing the mean square difference between Eq.~(\ref{eq:P_Dchi}) and the PDF from the mock catalogs.  We have found that fitting the PDF and CDF makes a negligible difference.  We put small offsets on $x$ values to avoid the overlapping of points.  To calculate the error bars, we divide the Uchuu simulation into 250 subvolumes, repeat the calculations independently for each volume, and calculate the standard deviation of the resulting parameters.  For each model, we show in the legend its deviation from the SDSS $\fproj$ using the reduced chi-squared $\chi_\nu^2 = (1/ \nu)\sum (\fproj^{\rm sim} - \fproj^{\rm SDSS})^2/\sigma_{\rm SDSS}^2 $, where $\nu=6$ is the number of richness bins.

Out mock catalogs reproduce the decreasing $\fproj$ with richness in \Myles.  The overall amplitude of $\fproj$ is sensitive to the projection depths.  We find that $\qproj$ = 180 $\hiMpc$ and $\sigproj$ = 100 $\hiMpc$ agree with the \Myles results.  As described in Sec.~\ref{sec:pmem_models}, these two models have approximately the same effective cylinder volume.    These models slightly overpredict $\fproj$ in the high-richness bin, but the statistical uncertainty in that bin is also larger.

Figure~\ref{fig:sigma_cl} shows the best-fit $\sigcl$ parameters for the Gaussian model with $\sigproj = $ 100, 110, 120 $\hiMpc$.  The $\sigcl$ values reflect the velocity dispersion of galaxies in clusters and are thus insensitive to the projection kernel.  Our $\sigcl$ values are consistent with \Myles and \Rozo for high-richness clusters but are slightly higher for low-richness clusters.  That is, our satellite velocity assignment based on dark matter particle velocities slightly overestimates the velocities of galaxies in low-richness clusters.

\subsection{Richness remeasured on a redshift grid}

Following \Costanzi, we use $\lambda(z)$ from the SDSS \redmapper clusters, measured as follows. For a given \redmapper cluster catalog and its associated red-sequence model, we offset a cluster's redshift $\zcl$ by $\Dz$ and recalculate the richness based on the red-sequence model at $z = \zcl + \Dz$.  We repeat this procedure using a range of $\Dz$ to construct $\lambda(z)$.
At $\zcl$, we recover the fiducial $\lambda$ from \redmapper. 
As $|\Dz|$ increases, fewer cluster members are consistent with the red-sequence model at $z=\zcl+\Dz$, and $\lambda(z)$ becomes lower.  The width of the projection depth determines how fast $\lambda(z)$ decreases.  For a wider projection depth, we expect that $\lambda$ would decrease slowly with $\Dz$.   
We restack the $\lambda(z)$ curves measured in \Costanzi; while \Costanzi focuses on the clean lines of sight [clusters with the 5\% narrowest $\lambda(z)$], we stack \text{all} clusters with $\lambda > 20$.  As a result, our stacked $\lambda(z)$ is broader than that in \Costanzi.

To mimic the SDSS $\lambda(z)$ measurement using our mock catalogs, we similarly offset the cluster center along the line of sight on a grid, from $-200$ to $200~\hiMpc$ with intervals of $10~\hiMpc$, and recalculate the richness $\lambda(\Dchi)$.  We rescale the curves so that $\lambda(\Dchi=0) = 1$ and stack the curves of all clusters.  We have verified that this $\lambda(\Dchi)$ recovers with the shape of $\pproj(\Dchi)$ for calculating counts in cylinders.

Figure~\ref{fig:lambda_z} shows the $\lambda(\Dchi)$ calculated from our mock catalogs compared with that from SDSS. 
The colored curves and bands correspond to the mean and standard deviation of $\lambda(\Dchi)$ from each mock catalog.
In the legend, we show the reduced chi-squared $\chi_\nu^2 = 
(1/ \nu)\sum_i (\lambda_{{\rm sim},i} - \lambda_{{\rm SDSS},i})^2/\sigma^2_{{\rm sim},i}$, where $i$ runs through the redshift grid points, and $\nu$ is the number of these points.  The left-hand panel shows the quadratic model, while the right-hand panel shows the Gaussian model.  We do not show the uniform model because it has a much larger deviation ($\chi_\nu^2 \gtrsim 2$) due to its sharp edges.
The SDSS result favors a quadratic kernel of $\qproj=150~\hiMpc$, which is slightly smaller than the value preferred by $\fproj$ ($\qproj=180~\hiMpc$) presented in the previous section.  The Gaussian kernel with $\sigproj=80~\hiMpc$ agrees reasonably well with the data, but it tends to be too narrow near the peak and too broad at the outskirts.

This $\qproj=150~\hiMpc$ derived from our sample is seemingly inconsistent with the main results presented in \Costanzi; their Eq.~(10) shows a projection width $\sz=0.034$ at $\zcl=0.1$, which corresponds to $90~\hiMpc$.  The apparent discrepancy is due to the different analysis choices: While we use all clusters with $\lambda > 20$, they use clusters with a clean line of sight.   In addition, while we fit the stacked $\lambda(z)$ curves, they fit one $\lambda(z)$ curve at a time and derive the $\sigma_z$ values for individual clusters, which tend to be smaller than the values derived from stacked $\lambda(z)$.

\section{Discussion}
\label{sec:discussion}

We have shown that cluster abundance and lensing signal are insensitive to projection depth, while the member redshift distribution and  $\lambda(z)$ are sensitive to the projection depth and can be used to constrain the projection kernel.  In this paper, we fix HOD and cosmological parameters and focus on the projection effects.  In our future work, we will explore the impact of HOD and cosmological parameters.

\citet{Sunayama23} uses the cross-correlation between spectroscopic galaxies and \redmapper clusters, $w_{\rm p, cg}$, to derive the projection depth.  The basic idea is that $w_{\rm p, cg}$ increases with the line-of-sight integration limit $\pi_{\rm max}$ until $\pi_{\rm max}$ reaches the projection depth.  They use a uniform cylinder model, and their best-fit projection depth is 136 $\hiMpc$.  This projection depth generally agrees with our projection kernel.

Our simulation framework will facilitate the analyses of the cross-correlation between clusters, galaxies, and shear \cite{Salcedo20, Zeng23}.  So far, most cluster and galaxy analyses are performed separately.  Cluster analyses (cluster counts, lensing, and clustering) tend to focus on mass calibration, while galaxy analyses (galaxy clustering and galaxy--galaxy lensing) tend to focus on calibrating the galaxy bias \cite{To21a}.  Our simulation-based framework presents an exciting possibility of unifying the analyses of clusters and galaxies because all summary statistics can be predicted self-consistently.  Further, if we design our cluster and galaxy samples based on the same color and magnitude selection, they can be described by the same set of HOD parameters.  In this way, the auto- and cross-correlation functions of clusters and galaxies can put tight constraints on HOD parameters.

The SDSS data used in this paper represents a rare occasion of 84\% complete spectroscopic cluster members for approximately 2000 clusters [see \cite{Wetzell21, Sohn18} for smaller samples].  In contrast, the spectroscopic sample from the Dark Energy Spectroscopic Instrument (DESI) tends to be highly incomplete in cluster regions; nevertheless, it might be possible to stack many clusters to obtain the average redshift distribution of clusters.  The Roman Space Telescope and \textit{Euclid} will both have galaxy redshifts from grism spectroscopy, which may be difficult to obtain for cluster galaxies because of the lack of emission lines and the crowding of cluster galaxies.  However, we can leverage the incomplete spectroscopic samples to study the redshift distribution of cluster members using an approach similar to clustering redshift.  For example, if we detect significant cross-correlation between cluster members and emission-line galaxies at different redshifts, it is an indication of projection effects.  Such an approach could be essential for future cluster studies combining photometric and spectroscopic data.

In this work, we focus on optical cluster observables.  In the future, we plan to model gas observables, such as x-ray luminosity/temperature and SZ signal, based on the same halo populations.  \citet{Zhou23} have shown that a joint selection between optical and gas observables can directly constrain the correlated scatter of richness and lensing caused by projection effects.   We plan to assign gas properties to either dark matter halos or particles \cite{Keruzore23, Zhou23, Keruzore24}.  This approach will allow for self-consistent modeling for projection effects.

\section{Summary}
\label{sec:summary}

We have presented a simple method for generating mock optical cluster catalogs, focusing on the projection effects associated with the redshift uncertainties of member galaxies.  We simulate red galaxies in halos in a wide range of mass using an HOD model.  To simulate the projection effects in the cluster-finding process, we use counts in cylinders to include galaxies along the cluster line of sight as projected members.  We use three parametrized models to describe the projection effects: uniform, quadratic, and Gaussian functions (Sec.~\ref{sec:pmem_models}).  We compare our model predictions with the data vectors measured from the SDSS \redmapper clusters: counts (Fig.~\ref{fig:abundance}), stacked lensing (Fig.~\ref{fig:lensing}), spectroscopic redshift distribution of member galaxies [$\mathcal{P}(z)$, Fig.~\ref{fig:delz_hist} and Fig.~\ref{fig:fproj}], and the richness remeasured on a redshift grid [$\lambda(z)$, Fig.~\ref{fig:lambda_z}].

We have found that, with the background galaxy number subtraction, the cluster counts vs.~richness are insensitive to the projection model.  Similarly, with abundance matching, the stacked cluster lensing signal is insensitive to the projection model.   The other data vectors, $\mathcal{P}(z)$ and $\lambda(z)$, are more sensitive to the projection kernel.  We have found that a quadratic projection kernel with a characteristic depth $\qproj \approx 180~\hiMpc$ ($\Dz \approx 0.06$, $v \approx 18,000~\kms$) closely reproduces the observed $P(z)$ and $\lambda(z)$.  
To compare the characteristic depths of different kernels, we convert them by matching the effective cylinder volume (Sec.~\ref{sec:pmem_models}).
A uniform projection kernel cannot reproduce the shapes of $\mathcal{P}(z)$ and $\lambda(z)$.

Our simple approach for generating mock catalogs for clusters provides a promising way for simulation-based inferences.  For example, for a given set of HOD and projection parameters, we can generate a cluster mock catalog and its associated data vectors, which can be compared with the observed data vectors to constrain cosmological and nuisance parameters \cite{Salcedo23}.  We expect that our approach can also be generalized to include multiwavelength observables by pasting gas properties to halos \cite{Keruzore23, Zhou23, Keruzore24}, enabling self-consistent analyses of galaxy-gas-halo connections.

\section*{Acknowledgements}

We thank the anonymous reviewer for helpful suggestions.
We also thank the Uchuu team for providing the $N$-body simulation suites.
A.L., H.-Y.W., S.C., and L.Y.~are supported by the DOE Award No.~DE-SC0021916.
A.L.~additionally receives Boise State's HERC Fellowship for Undergraduate Research.
A.N.S.~is supported by the DOE Awards No.~DE-SC0009913 and No.~DE-SC0020247.
M.C.~is supported by the PRIN 2022 project EMC2 --- Euclid Mission Cluster Cosmology: unlock the full cosmological utility of the Euclid photometric cluster catalog (code No.~J53D23001620006).
The computations in this paper have been performed on the R2 and Borah computer clusters\footnote{\href{https://scholarworks.boisestate.edu/oit/3/}{DOI: 10.18122/oit/3/boisestate}} provided by Boise State University's Research Computing Center.

\section*{Data Availability}

The data used in this work is available upon request.

\appendix

\section{Galaxy color vs.~distance uncertainties}
\label{app:color_dist}

\begin{figure}
\includegraphics[width=\columnwidth]{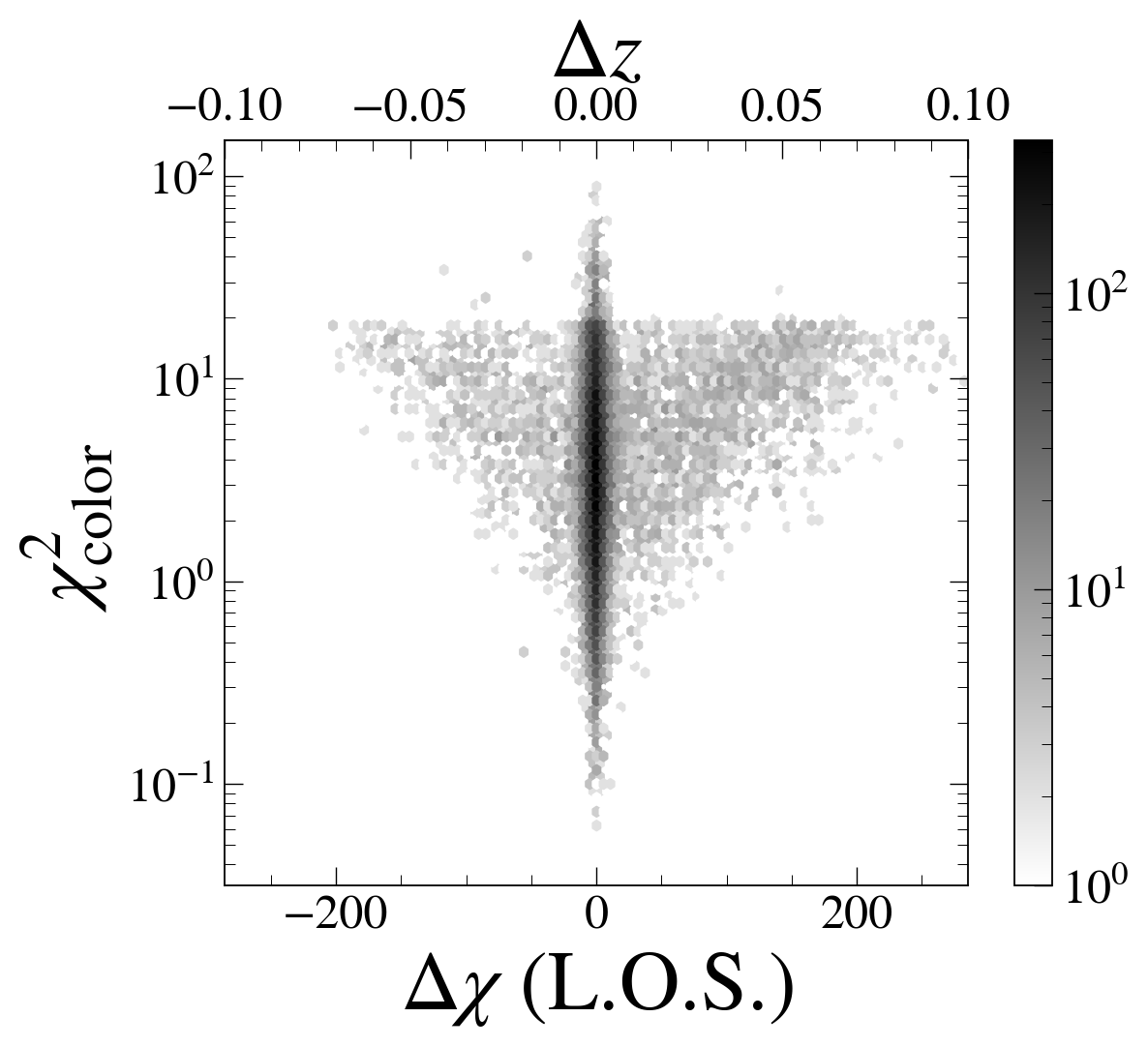}
\caption{SDSS \redmapper member galaxy color vs.~the line-of-sight comoving distance from the central galaxy, $\Dchi$.  The galaxy color is quantified by the deviation from the red-sequence model, $\chisqc$ [Eq.~(\ref{eq:chisqc})].  The \redmapper algorithm selects member galaxies with $\chisqc < 20$ and central galaxies with $\chisqc < 100$.  The dark region near $|\Dchi| = 0$ corresponds to the true members, and the wings correspond to projected members.  A larger $\chisqc$ corresponds to a wider range of line-of-sight distances.  A $\chisqc = 20$ corresponds to a distance upper limit of $\approx$ 200 $\hiMpc$.}
\label{fig:chisq_dist}
\end{figure}

When the \redmapper algorithm searches for cluster members, each galaxy's color $\bf{c}$ and magnitude $m$ are compared with the red-sequence model at a given redshift $z$.  The distance from the mean red-sequence color is quantified by 
\beq
\chisqc(z) = (\mathbf{c} - \avg{\mathbf{c} | z, m})^T \mathbf{C}^{-1}(\mathbf{c} - \avg{\mathbf{c} | z, m})
\label{eq:chisqc}
\eeq
[see Eq.~(6) in \cite{Rykoff14}], where $\mathbf{C}^{-1}$ is the covariance matrix of colors.  A larger $\chisqc$ at a given redshift can result from a scatter in color (intrinsic scatter or photometric noise) or from a large deviation from that redshift.  We would like to see how the color cut and the distance uncertainties are related to each other.

Figure~\ref{fig:chisq_dist} shows $\chisqc$ vs.~line-of-sight distance for the SDSS \redmapper members with spectroscopic redshifts (introduced in Sec.~\ref{sec:redmapper_catalog}).  We show $\Delta z = z_{\rm spec, mem} - \zCG$ on the top $x$ axis and the corresponding line-of-sight comoving distance ($\Dchi$) on the bottom $x$ axis.  The hexagonal binning shade corresponds to the logarithmic number of galaxies in each bin.

The upper edge of the distribution corresponds to \redmapper's color cut for members, $\chisqc < 20$.  The points above $\chisqc = 20$ correspond to central galaxies, which have a less stringent color cut, $\chisqc < 100$.  The dark area in the middle with $|\Dchi|\approx 0$ corresponds to the true members.  For $\chisqc < 1$, most galaxies have $|\Dchi|\approx 0$, indicating the small distance uncertainties for these galaxies.  As $\chisqc$ increases, the range of the $|\Dchi|$ also increases.  A $\chisqc < 20$ cut approximately corresponds to a distance uncertainty upper limit of 200 $\hiMpc$.  We note this distance uncertainty is the upper limit rather than the mean; most members have much smaller line-of-sight distances than this upper limit.   The fiducial redMaGiC sample has a more stringent cut $\chisqc \lesssim 2 $ and a typical redshift uncertainty of 0.017 \cite{Rozo16redMaGiC}, which is also consistent with this figure.

\begin{figure}
\includegraphics[width=\columnwidth]{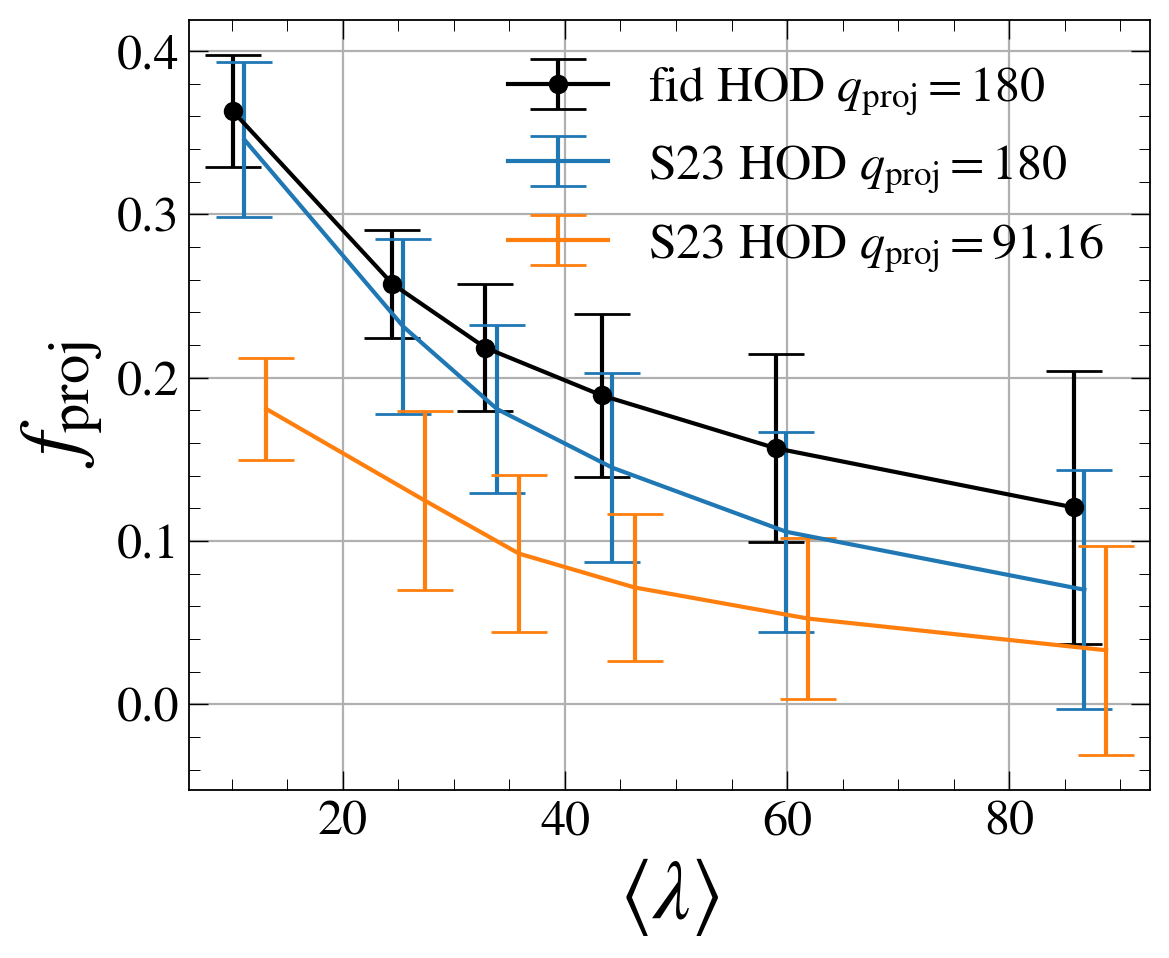}
\caption{Projected fraction $\fproj$ with an alternative HOD motivated by \protect\citet{Salcedo23}, which produces richnesses lower than our fiducial HOD.  We show our fiducial HOD with $\qproj=180~\hiMpc$ (black, also presented in  Fig.~\ref{fig:fproj}), the alternative HOD with $\qproj=180~\hiMpc$ (blue), and the alternative HOD with $\qproj=91.16~\hiMpc$ (orange).  The projected fraction is relatively insensitive to HOD parameters but is sensitive to the projection depth.
}
\label{fig:salcedo_hod}
\end{figure}
\section{Testing an alternative HOD}
\label{app:alt_HOD}

In the main text, we fix the HOD parameters to the fiducial values in \citet{Sunayama20}.  In this appendix, we test an alternative HOD model motivated by \citet{Salcedo23}:
$\alpha=1.22$,
$\log_{10}M_1 = 13.67$ (corresponding to $\log_{10}M_{20}=14.75$), 
$\log_{10}M_0 = 11.7$,
$\log_{10}M_{\rm min} = 12.48$, and
$\sigma_{\log M} = 0.1$.
This model gives a lower richness compared with our fiducial model.  Our calculation is slightly different from \cite{Salcedo23}: We use $z=0.1$ instead of $z=0.3$, and we use Uchuu instead of AbacusSummit.  Given that no observational data of $\fproj$ is available at $z=0.3$, we focus on the $z=0.1$ comparison.

Figure~\ref{fig:salcedo_hod} shows the projection fraction $\fproj$ from the alternative HOD, with a quadratic projection model with two different projection depths: $\qproj=91.76~\hiMpc$ (motivated by \Costanzi) and $\qproj = 180~\hiMpc$ (motivated by our results in Sec.~\ref{sec:Pz}).  The latter predicts a $\fproj$ similar to the fiducial HOD (black), while the former predicts a low projected fraction.  Comparing this result with Fig.~\ref{fig:fproj}, we can see that $\fproj$ is sensitive to the projection depth but relatively insensitive to the HOD parameters.  In future work, we plan to investigate the impact of HOD parameters on various data vectors.

\bibliographystyle{apsrev4-2-author-truncate}
\bibliography{master_refs}

\end{document}